# Carrier Transport in High Mobility InAs Nanowire Junctionless Transistors

*Aniruddha Konar*[*,†], *John Mathew*[⊥], *Kaushik. Nayak*[ǂ,‡], *Mohit. Bajaj*[†], *Rajan K. Pandey*[†], *Sajal Dhara*[⊥], *K. V. R. M. Murali*[†,‡], *Mandar Deshmukh*[⊥]

[†]IBM Semiconductor Research and Development Center, Manyata Embassy Business Park, Nagawara, Bangalore – 560 045, India

[ǂ] Department of Electrical Engineering, Indian Institute of Technology Bombay, Powai, Mumbai – 400 076, India

[⊥] Department of Condensed Matter Physics and Materials Science, Tata Institute of Fundamental Research, 1 Homi Bhabha Road, Mumbai 400005

[*]E-mail: (A. K.) anikonar@in.ibm.com

Ability to understand and model the performance limits of nanowire transistors is the key to design of next generation devices. Here, we report studies on high-mobility junction-less gate-all-around nanowire field effect transistor with carrier mobility reaching 2000 cm$^2$/V.s at room temperature. Temperature-dependent transport measurements reveal activated transport at low temperatures due to surface donors, while at room temperature the transport shows a diffusive behavior. From the conductivity data, the extracted value of sound velocity in InAs nanowires is found to be an order less than the bulk. This low sound velocity is attributed to the extended crystal defects that ubiquitously appear in these nanowires. Analyzing the temperature-dependent mobility data, we identify the key scattering mechanisms limiting the carrier transport in these nanowires. Finally, using these scattering models, we perform drift-diffusion based transport simulations of a nanowire field-effect transistor and compare the device performances with experimental measurements. Our device modeling provides insight into performance limits of InAs nanowire transistors and can be used as a predictive methodology for nanowire-based integrated circuits.

Field effect transistor is the building block of integrated circuits and is key to new technologies. The aggressive scaling has pushed the silicon-based planar Metal-Oxide Semiconductor Field Effect Transistor (MOSFET) technology to the point where transistor performances cannot be enhanced by simply reducing dimensions. For further scaling of the transistor, several [1] alternative materials (other than silicon) and device geometries have been investigated including the Fin-Field

---

[‡] Currently with GlobalFoundaries

Effect Transistor [2,3], Tri-gate [4], Omega gate [5], and Gate All-Around or wrap gate (GAA) devices[6]. Though the fabrication of GAA device geometry is much more complex in a top-down approach, GAA (wrap gate) devices offer higher performance due to its superior electrostatic controls compared to other geometries. In this work, we fabricate and evaluate the performance of high mobility InAs nanowire junctionless transistors (JLT), synthesised through a combination of a simpler bottom-up approach and conventional lithographic techniques [7].

Although, InAs nanowire devices [8–11] have higher mobility than silicon based devices, the mobility in nanowires is significantly lower than that of bulk InAs[7–9,12] due to a variety of factors including increased electron/hole scattering rates and surface scattering mechanisms [8]. A detailed understanding of various carrier scattering mechanisms is essential to be able to improve experimental methods to build higher performance nanowire transistors in the future. In this work, we develop microscopic models to understand the experimental data and the limitations of the performance of the fabricated InAs GAA junctionless InAs nanowire transistors. To do so, we first identify the important scattering phenomena limiting the carrier transport in these nanowires. Using identified scattering mechanisms, we perform transport simulation in a drift-diffusion based device simulation environment FIELDAY[13], to characterize and evaluate performance limitation of InAs GAA JLT for possible applications in large-scale circuits.

InAs nanowires were grown via the Vapor Liquid Solid (VLS) technique using MOCVD (metal-organic chemical vapor deposition), as described in our previous work [7]. Nevertheless, for the sake of completeness, we pictorially describe the process flow for device fabrication in Figure 1(a). Details of the device fabrication is described in the supplementary section. After all the lithographic steps, the final NWFET with source, drain and gate contacts is shown schematically in Fig.1 (b) as well as the scanning microscope image in Fig. 1 (c). Transport measurements are carried out in a helium flow cryostat using a lock-in technique as well as a DC measurement, where the drain current ($I_d$) or conductance is measured as a function of gate voltage ($V_g$). Mobility can be extracted from the gating curves by using the transconductance of the device $\mu_{FE} = (dG/dV_g)(L^2/C)$, where $\mu_{FE}$ is the field effect mobility, $dG/dV_g$ is the transconductance, $L$ is the length of the channel, and $C$ is the gate capacitance. In our bottom-up approach of fabricating NWFETs, source, drain and the gate contacts are not self-aligned. This usually leads to un-gated regions (known as access region) between the source/drain contact and the gate contact as shown in Fig. 1 (c). In our mobility calculations, we carefully remove the contribution of the resistance coming from the access regions (see supplementary section). Hence the conductance and mobility reported here should be attributed to the gated part of the nanowire only.

Figure 2 (a) shows the conductance (gated part only) of InAs junctionless nanowire field effect transistors (NWFETs) as a function of temperature for various gate voltages. We note that the conductance is decreasing with decreasing $V_g$. This is due to the carrier depletion from the nanowire channel at lower $V_g$. For a particular $V_g$ (carrier density), the conductance first increases with temperature (as shown in Figure 2 (b)) up to a critical temperature (~*140 K*) and then follow a rather complicated (first decrease and then increase) dependency on temperature. The increase of conductance with temperature is a typical signature of the donor activation process with temperature. At low temperatures, the donors are frozen and conductivity decreases with decreasing temperatures. We define nanowire conductance $G(T) = \sigma(T)\frac{A}{L}$, where $\sigma(T)$ is the carrier conductivity, $A$ is the cross section and $L$ is the length of the nanowire. Since conductance is proportional to conductivity; we will use conductance and conductivity synonymously in the rest of the paper. As nanowires are long ($L\sim 3$ μm), we can safely assume that the transport is diffusive in nature. This allows us to write temperature dependent conductivity $\sigma(T) = \frac{n(T)e^2 \tau(T)}{m^*}$, where $n(T)$ is the carrier density as a function of the temperature in the InAs channel, $e$ is the electronic charge, $m^*$ is the carrier effective mass, and $\tau(T)$ is the temperature dependent average momentum relaxation time. The dependence of the relaxation time on the temperature $T$ is determined by the dominant type of scattering mechanism. At low temperatures, phonons are frozen and electron-phonon scattering is less probable. Moreover, measurements show that (see Fig. 2(c)) the conductance increases with $V_g$ (carrier density). This conductivity enhancement is a typical signature of the dominance of charged impurity scattering, where conductivity increases (scattering time increases) with carrier density (~ Fermi energy) due to enhanced free carrier screening of the Coulomb (scattering) potential. Shift of threshold voltage at different temperature as shown in Fig. 2(c), confirms activation of carriers with temperatures along with the role of ionized dopants in the carrier transport. After the critical temperature, the complicated behavior of the conductivity is the reflection of the complex temperature dependence of the product $n(T)\tau(T)$ due to interplay of charged impurity and phonon scattering.

To investigate how the charged impurity scattering depends on the free carrier densities, we evaluate momentum relaxation time using Fermi's golden rule[14]. In our experiments, nanowires are thick (diameter d ~ 70 nm) and we assume a 2D carrier gas in the channel like planar field effect transistors (FETs). We assume that charged impurities are located at the nanowire-high-k interface. From our calculations, we find (see supplementary information for detail calculations) that the carrier conductivity has a quadratic dependence on carrier density as $\sigma(n) \sim n^2$. The carrier density in the channel at any gate-voltage has two parts, $n = n_e(V_g) + n_0 e^{-\frac{E_a}{k_B T}}$, where $n_e(V_g)$ is the field induced carrier density, $n_0$ is the total donor concentration, $E_a$ is the donor activation energy, $k_B$ is the Boltzmann constant and $T$ is the temperature. Fitting the low temperature (<150 K) conductivity data with a quadratic function of carrier density (see supplementary section), the activation energy $E_a$ can be extracted. Figure 3 (a) shows the experimental data along with the fitting with the above expression for three different gate voltages. With this conductivity expression, we were able to fit conductivity data for a wide range of gate voltage with increasing activation energy. Figure 3(b) shows the change of activation energy (barrier height) as a function of gate voltage as extracted from the conductance-temperature fit. This change of barrier height with gate voltage can stem out from two possible reasons; (i) deep donor states situated far above in the conduction band (CB) of InAs[15], (ii) effective barrier[16] coming from the CB offset of two different crystal phases of InAs at the twin boundary (see supplementary information) present in these nanowires. Our nanowires contain twin boundaries along the growth (transport) direction as shown in (Fig. S1) the supplementary information. A twin boundary can be perceived as a smooth interface (without any dangling bonds) between wurtzite (WZ) and the zincblende (ZB) crystal phase of InAs as shown in Fig. 3 (d). Differences in electronic band structures of WZ and ZB phase of InAs lead to effective barriers for carrier transport at the twin boundary (see Fig. 3(e). Figure 3 (c) shows the band lineup between WZ and ZB phase of InAs. It is clearly shown that there is 0.086 eV[17] of CB offset between two crystal phases of InAs. Whether the transport is tunneling dominated can be inferred from the transmission coefficient through this barrier. Approximating the barrier as a delta-function potential[18] as shown in Fig. 3(f), we write the strength of the delta-potential[19] barrier as $S = d\Delta E_C$, where $d$ is the twin boundary width (~few nm) and $\Delta E_C$ is the CB offset of the twin boundary. The transmission coefficient of carriers through this delta-function barrier is calculated as[20] $Tr(E_f) = \left[1 + \frac{m^* S^2}{2\hbar^2 E_f}\right]^{-1}$, where $m^*$ is the carrier effective mass and $\hbar$ is the reduced Planck's constant. For a few nm (2-3 nm) of barrier width and the band offset, we found that the barrier is almost transparent (Tr>0.6) over a wide range of electron densities (n>$10^{11}$/cm$^2$) – typical carrier densities present in the channel in the "ON" state of a transistor. The transparency of the barrier coming from CB offset indicates that nominal variation in the activation energy coming from deep level donors present in the CB of the InAs. Moreover, the thin transparent tunnel barrier at the WZ/ZB interface also validates the use of diffusive transport formalism in this work. To further investigate the diffusive transport through these NWFETs, we identify the dominant scattering mechanisms, and then extract relevant transport parameters from the mobility measurements. Figure 4 (a) shows the carrier mobility of InAs nanowire as a function of temperature. It should be noted that the mobility first increases and then falls drastically with temperatures. This can be explained by the dependence of scattering time with temperatures. For diffusive transport the carrier mobility is written as $\mu = \frac{e\tau(n)}{m^*}$. At low temperatures, the mobility increases with temperature. This is due to the fact that carrier density in the channel increases because of donor activation, with temperature, which leads to enhanced free carrier screening for Columbic perturbation and increases mobility.

For Columbic scattering, it can be shown (see supplementary information) that the scattering time $\tau_c(n) \sim k_F^2 \sim n$. As a result, we fit the low temperature mobility data with impurity-limited mobility model which has an exponential dependence on the temperature (see supplementary material for detail explanation). As mobility is measured at a carrier density of $n_{3D}$=3 x 10$^{17}$/cm$^3$, the corresponding equivalent 2D carrier density is calculated to be $n(V_g) \approx (n_{3D})^{2/3} \approx 4.48$ x 10$^{11}$/cm$^2$. With this $n(V_g)$ and extracted activation energy E$_a$~ 4.4 meV (from the conductivity data), low temperature mobility data calibration is shown in Fig. 4 (b). From the mobility fit, we extract the impurity density at the high-k/InAs interface is n$_{imp}$~1.32 x 10$^{13}$/cm$^2$, which is similar to the reported values in the literature[21,22]. For numerical calculations, we use effective mass of electrons m$^*$=0.023 m$_0$[5] (m$_0$=9.1x10$^{-31}$Kg is the free electron rest mass), and dielectric constant of InAs $\kappa_s = 15.15$. The donor density is extracted to be n$_0$~0.19 x 10$^{11}$/cm$^2$. This low donor density compared to the n(V$_g$) is justified by the fact that the mobility increases nominally with temperature (nominal increase of screening by thermally activated carriers in addition to the screening by electro-statically induced carriers by gate voltage). The degradation of carrier mobility at high temperatures (after the critical temperature ~100 K) indicates that some other scattering mechanism (in addition to the impurity scattering) is dominant at those elevated temperatures, as shown in Figure 4(a). At any temperature, in the presence of multiple scattering mechanisms, total carrier mobility in InAs nanowires can be written using Mathiessen's rule as $\mu_T^{-1} = \mu_{cou}^{-1} + \mu_{other}^{-1}$, where $\mu_{other}$ is the other dominant scattering mechanism at elevated temperatures. The mobility $\mu_{other}$ can be determined by subtracting Coulomb-limited mobility from the experimentally measure mobility, i.e. $\mu_{other}^{-1} = \mu_{exp}^{-1} - \mu_{Cou}^{-1}$. The extracted inverse mobility (~resistivity) $\mu_{other}^{-1}$ is plotted in Fig. 5 (a), as a function of temperature. The linear dependence of resistivity on temperature at these elevated temperatures can act as a guideline for determining the dominant scattering mechanism. The enhancement of resistivity with temperature is a typical signature of contribution coming from electron-phonon scattering. High energy polar optical phonons (phonon energy of $\hbar\omega_o = 30$ meV) of InAs are not quite activated at the temperature range (thermal energy of 12-22 meV) we are interested in. Detail calculations (see the supplementary section) reveal that the electron-acoustic phonon interaction limited resistivity of the 2DEG of InAs is linear with temperature. As a result, we fit the measured high-temperature (>150K) mobility data with electron-acoustic phonon interaction. The linear fit of the experimental data with acoustic-phonon scattering is shown in Fig.5 (b). From the slope of the fit we extract, we extract the deformation potential ($D_a$) to the sound velocity ($v_s$) ratio, $D_a/v_s$ ~ 0.026 (unit of eV.s/m), much higher than it bulk value (0.0025). Using the deformation potential close to bulk value $D_a$~10 eV, we found the sound velocity in our InAs nanowires $v_s$ ~ 0.5 km/s, which is almost one order lower than the longitudinal acoustic phonon velocity in bulk (2-3 km/s). This low phonon velocity can be attributed to the extended crystal defects, which results in average sound velocity lower than the bulk counterpart of InAs. It is known that the III-V (InAs and others) nanowires have a tendency to form twins [23–25] when the diameter of the nanowires exceeds a critical diameter ( see supporting information for a TEM image). These structural defect regions act as potential scattering centres for sound propagation thus reducing the average sound velocity lower than otherwise defect-free bulk InAs. This lower sound velocity results in stronger electro-acoustic phonon coupling which in turn degrades carrier mobility in these nanowires. In defect-free InAs nanowires, the electron-acoustic phonon scattering will be weaker (almost by an order of magnitude) than the electron-impurity scattering (assuming same areal impurity density similar to here) - and the mobility will be limited by the random ionized impurity arising due to fabrication process at the nanowire-high-k interface. Further optimization of growth and fabrication is expected to reduce the interfacial impurity density- this will reduce the Coulomb scattering and enhance the carrier mobility. Nevertheless, the InAs room temperature mobility reported here is superior to the electron mobility in bulk/nanowire Si (~1200/300 cm$^2$/V.s)[26] or Si-Ge core-shell hetero-structure nanowires[27] and comparable to GaN nanowire-based high electron mobility transistors (HEMTs)[28].

Having determined scattering mechanisms responsible for electronic transport through these nanowires, we perform extensive numerical simulation (details are in the supplementary section) to validate the extracted scattering/mobility parameters by characterizing the current-voltage measurement of NWFETs. To do so, we particularly use acoustic phonon and remote Coulomb scattering mobility model in the drift-diffusion transport equation with parameters extracted in the previous sections. The 3-D transport simulations of the InAs GAA n-NWFET are carried out using in-house device simulation program FIELDAY[13,13,29]. We consider cylindrical InAs nanowire FET with HfO$_2$ gate oxide and metal gate. To compare with the experimental data, the device parameters considered by us are similar to the lithographically fabricated NWFETs, i.e. nanowire diameter ($d_{NW}$) = 70 nm, gate length ($L_G$) = 2 µm, and gate oxide thickness ($T_{OX}$) = 10 nm, and body doping (N$_D$=1.5 x 10$^{18}$/cm$^3$). The gate unwrapped region boundary surfaces of the InAs nanowire are kept electrically insulated. A series resistance (R$_{ext}$ = 38 KΩ ) due to the un-gated region is used along with the tuned electrostatics and mobility model parameters to match the simulation predictions with the experimental transfer characteristics at various V$_{DS}$ values) from 20 mV to 80 mV in steps of 20 mV. In our simulation, we consider electron-acoustic phonon scattering, impurity scattering, remote coulomb scattering due to fixed charges in gate oxide, and surface roughness scattering-limited mobility models for carriers in the InAs. Figure 5 (b), shows the simulated transfer characteristics of the InAs n-NWFET in comparison with the experimental data. In our simulation, we used remote impurity density of $n_{imp} \sim 1.32 \times 10^{13}/cm^2$ along with sound velocity $v_s = 0.5$ km/s as extracted in the earlier section from carrier mobility analysis. Further the drift-diffusion transport related parameters (carrier scattering parameters, carrier saturation velocity, and lumped source/drain series resistance) are tuned to match the experimental transfer characteristics. The excellent agreement between the simulated and measured device characteristics confirms the accuracy of carrier scattering analysis presented in previous sections. These set of mobility parameters can be used to predict the performance of a large scale NWFET-based circuits.

In conclusion, we have fabricated high-mobility junctionless gate all around InAs nanowires transistors showing room temperature carrier mobility ~ 2200 cm$^2$/V.s. Our device modelling confirms that the transport in these VLS-grown InAs nanowire field effect transistors are limited by the presence of extended structural defects (twin boundaries) that are conjectured to affect the electrical properties[16,24]. Based on measured transport data, we develop carrier scattering models which are subsequently used in a device simulator to predict the device performance of these NWFETs. At low temperatures, measurement shows an activated transport in NWFETs due to donors present in the bulk as well as at the nanowire-high-k interface. At room temperature, the carrier transport in our NWFETs is limited by the interplay of remote Coulomb scattering and acoustic phonon scattering. In addition, the extended structural defects (twin boundaries) significantly degrade the performance of the device at room temperature by lowering the average phonon velocity thus enhancing the electron-acoustic phonon scattering rate. One of the strategies that will result in significantly improved device performance includes passivation to eliminate interfacial defects as well as removal/minimizing structural defects like twin boundaries via improved controlled[25] growth mechanism.

ASSOCIATE CONTENTS

The supporting information contains details of the device fabrication, data analysis, carrier scatterings as well as device modeling.

ACKNOWLEDGMENT
Mandar Deshmukh acknowledges the support from Government of India and IBM through the IBM faculty award. We acknowledge Prof.Arnab Bhattacharya and Mahesh Gokhale for providing the InAs nanowires used in this study.

**FIGURE CAPTIONS**

**Figure 1**. Wrap gate all around device fabrication and structure. (a) The fabrication of the wrap gate transistor starts by depositing InAs nanowires in a sandwich of polymer resist. Sequence of polymer layer resist was used to ensure a large degree of undercut to enable a clean lift-off process.

Using e-beam lithography a region is patterned to define the gate. Following development atomic layer deposition (ALD) is used to deposit $HfO_2$ that is the gate dielectric. Following this the drain source electrodes are patterned; this is followed by development. The final step is deposition of electrodes subsequent to an in-situ plasma cleaning to remove residue and amorphous oxide. Lift-off process gives the wrap-gate transistor. (b) InAs Nanowire field-effect transistor-schematic showing junctionless transistor. (c) False colored scanning electron microscope (SEM) image of the device with the source (S), drain (D) and gate (G) indicated.

**Figure 2.** Conductance measurements as a function of temperature. (a) Experimental conductance vs temperature data for various gate voltages, (b) enhancement of conductance plotted up to the critical temperature for various gate voltages (inset shows the typical conductivity variation below threshold voltage). We define critical temperature up to the temperature where conductance (conductivity) increases, (c) conductance as function of gate voltages, note that the threshold voltage shifts at various temperature due to activation of donor states.

**Figure 3.** (a) Measured conductance fitted with activation formula for three different gate voltages. (b) Change of activation energy as a function of gate voltage almost upto pinch-off state. This increase is attributed to the deep donor state deep in the conduction band in the InAs nanowires- effective activation barrier changes with Fermi level movement with gate voltages. (c) Band alignment of WZ and ZB phase of InAs, (d) schematic of twin defects in our InAs nanowires, (e) schematic "abrupt" barrier at the WZ/ZB interface, and (f) approximation of CB offset barrier as Dirac-delta potential. In the "ON" state of the NWFET, we found that barrier due to CB offset is almost transparent to the electrons thus making transport diffusive.

**Figure 4**. Comparison with theoretical modeling: (a) carrier mobility as a function of temperature of InAs nanowire FET; at low temperature mobility is impurity scattering limited and at high

temperatures phonon scattering dominates over other scattering mechanism, and (b) fit of low temperature carrier mobility with the impurity-limited mobility model. Activation of carriers with temperature is taken account in the impurity mobility model.

**Figure 5.** Understanding the InAs gate all around FET response: (a) fit of the high-temperature mobility data with phonon scattering mechanism. Extracted sound velocity is found to be almost an order lower due to extended twin boundaries in our InAs nanowires. (b) Comparison of simulated InAs depletion n-NWFET transfer characteristics with experimental data. We use impurity and phonon scattering model in our in-house 3D self-consistent drift-diffusion (with quantum correction) device simulator for transfer characteristics of InAs nanowires. An excellent agreement is found between simulation results and measurement data corroborating our scattering models for diffusive transport developed in this work.

.

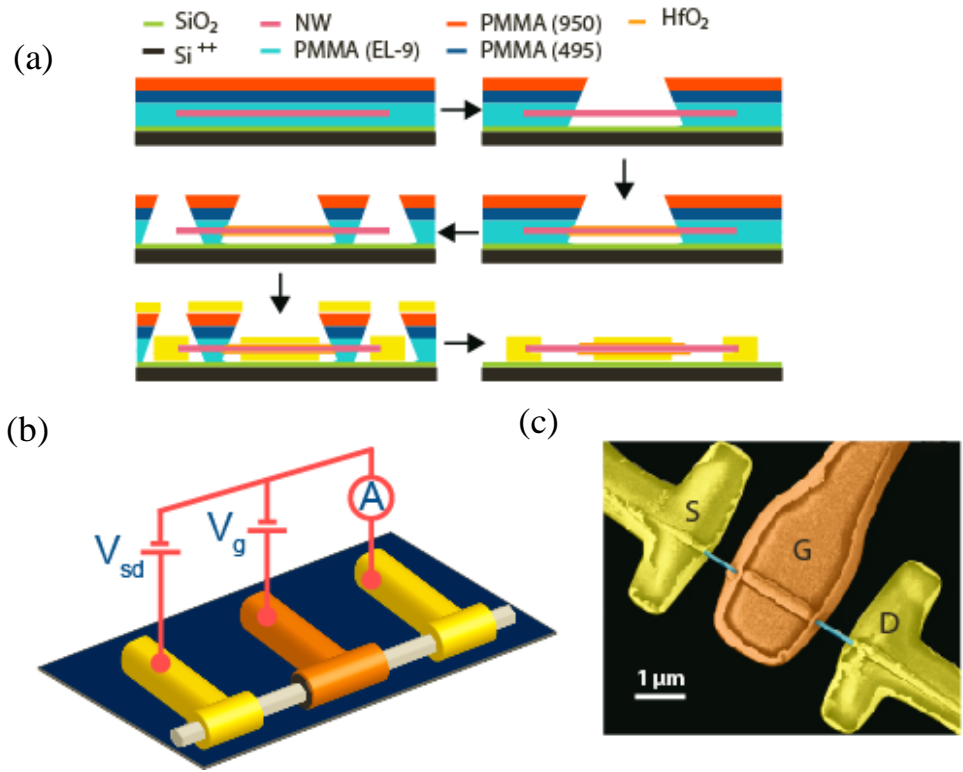

Figure 1. Wrap gate all around device fabrication and structure. (a) The fabrication of the wrap gate transistor starts by depositing InAs nanowires in a sandwich of polymer resist. Sequence of polymer layer resist was used to ensure a large degree of undercut to enable a clean lift-off process. Using e-beam lithography a region is patterned to define the gate. Following development atomic layer deposition (ALD) is used to deposit $HfO_2$ that is the gate dielectric. Following this the drain source electrodes are patterned; this is followed by development. The final step is deposition of electrodes subsequent to in-situ plasma cleaning to remove residue and amorphous oxide. Lift-off process gives the wrap-gate transistor. (b) InAs Nanowire field-effect transistor-schematic showing junctionless transistor. (c) False colored scanning electron microscope (SEM) image of the device with the source (S), drain (D) and gate (G) indicated.

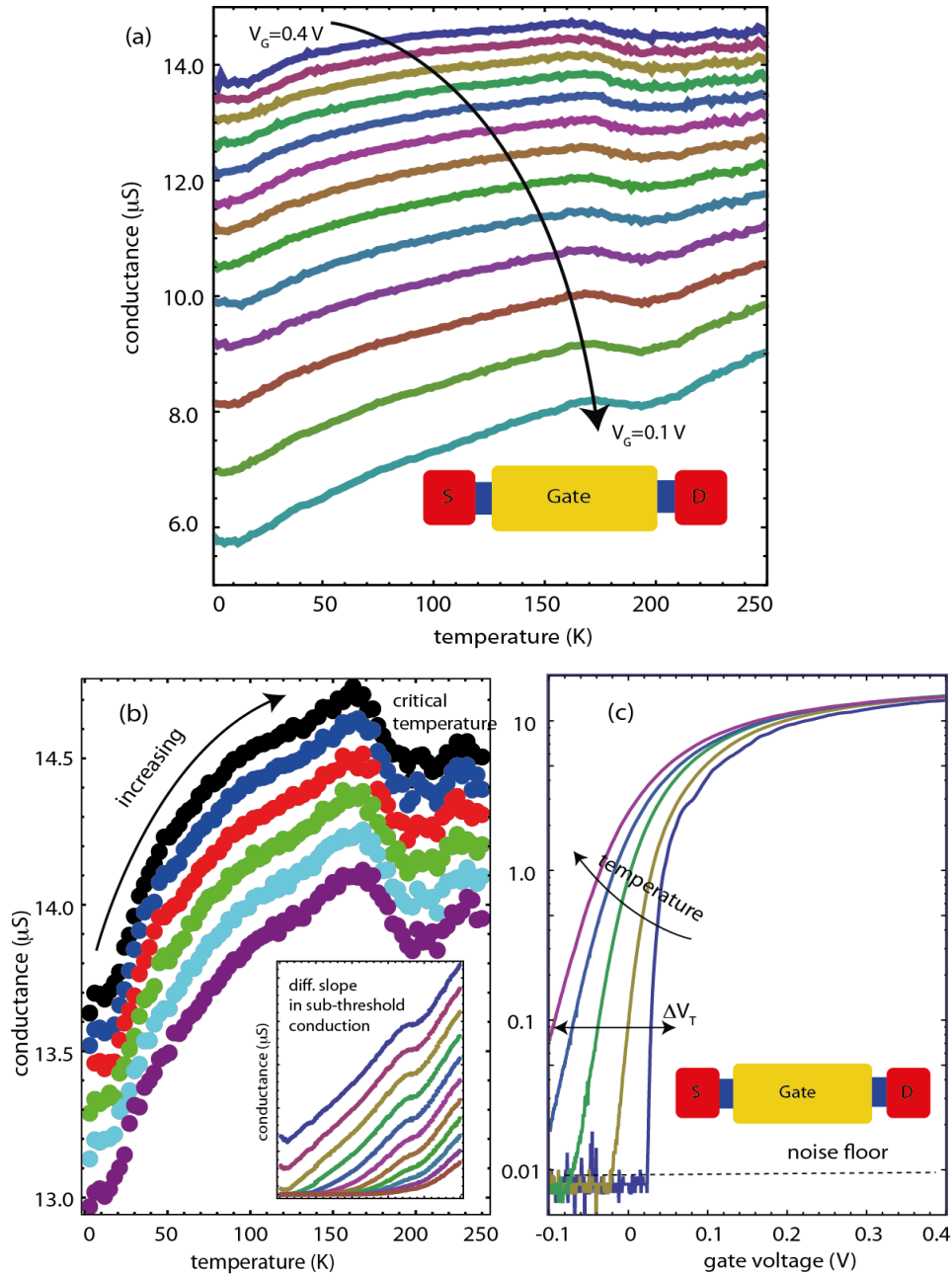

**Figure 2**. Conductance measurements as a function of temperature. (a) Experimental Conductance vs temperature data for various gate voltages, (b) enhancement of conductance plotted up to the critical temperature for various gate voltages (inset shows the typical conductivity variation below threshold voltage). We define critical temperature up to the temperature where conductance (conductivity) increases, (c) conductance as function of gate voltages- please note the threshold voltage shifts at various temperature due to activation of donor states.

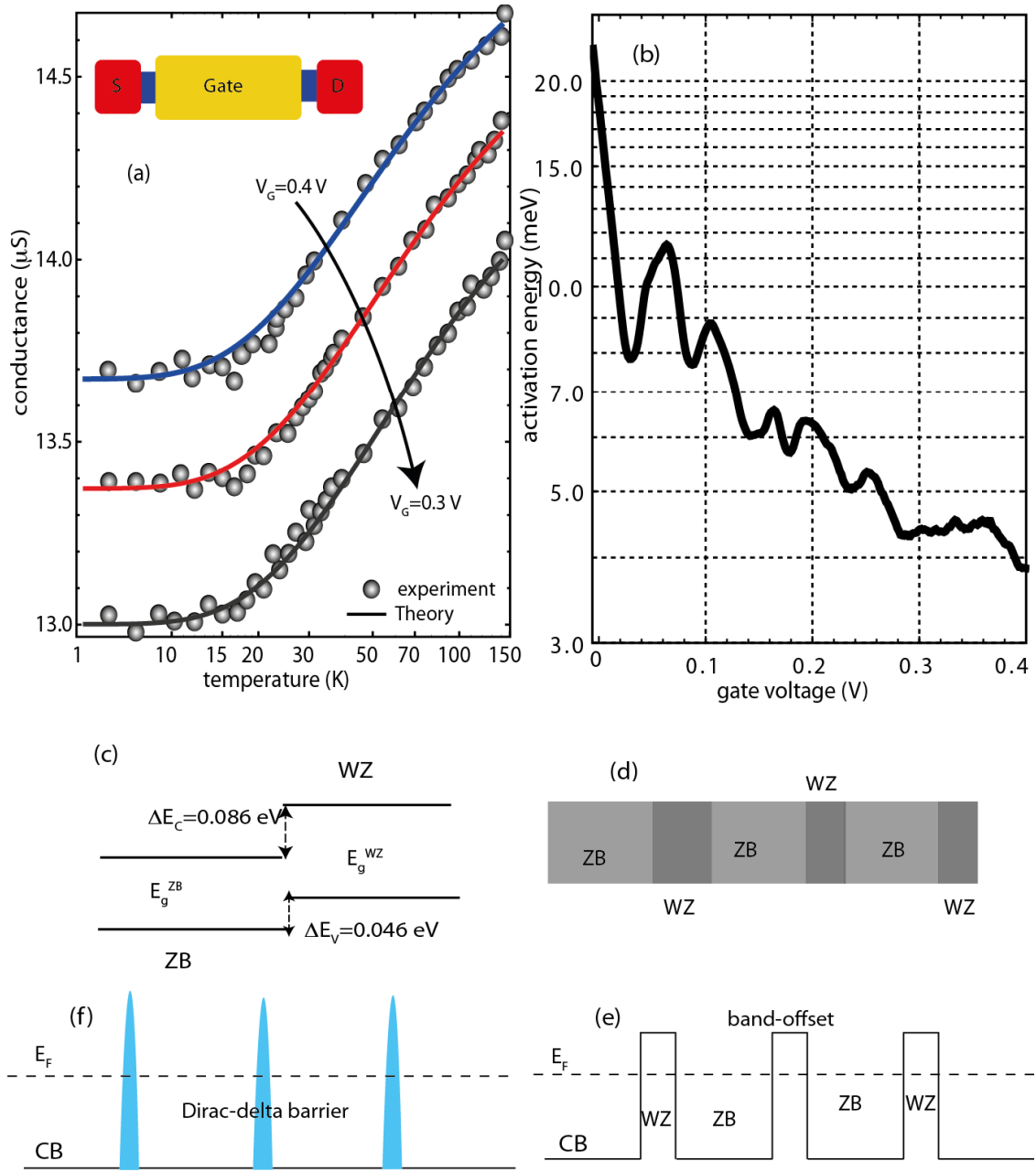

**Figure 3**. (a) Measured conductance fitted with activation formula (see main text) for three different gate voltages. (b) Change of activation energy as a function of gate voltage almost up to pinch-off state. This increase is attributed to the deep donor state deep in the conduction band in the InAs nanowires- effective activation barrier changes with Fermi level movement with gate voltages. (c) Band alignment of WZ and ZB phase of InAs, (d) schematic of twin defects in our InAs nanowires, (e) schematic "abrupt" barrier at the WZ/ZB interface, and (f) approximation of CB offset barrier as Dirac-delta potential. In the "ON" state of the NWFET, we found that barrier due to CB offset is almost transparent to the electrons thus making transport diffusive.

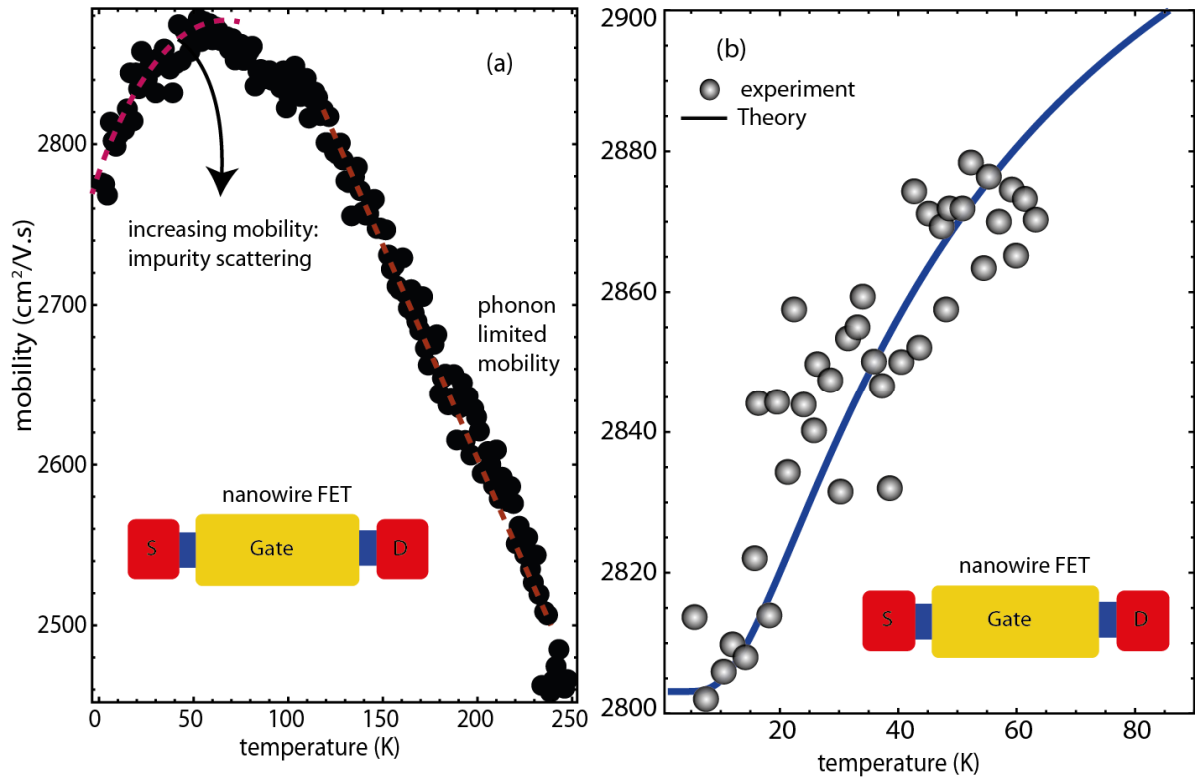

**Figure 4.** Comparison with theoretical modelling: (a) carrier mobility as a function of temperature of InAs nanowire FET; at low temperature mobility is impurity scattering limited and at high temperatures phonon scattering dominates over other scattering mechanism, and (b) fit of low temperature carrier mobility with the impurity-limited mobility model. Activation of carriers with temperature is taken account in the impurity mobility model.

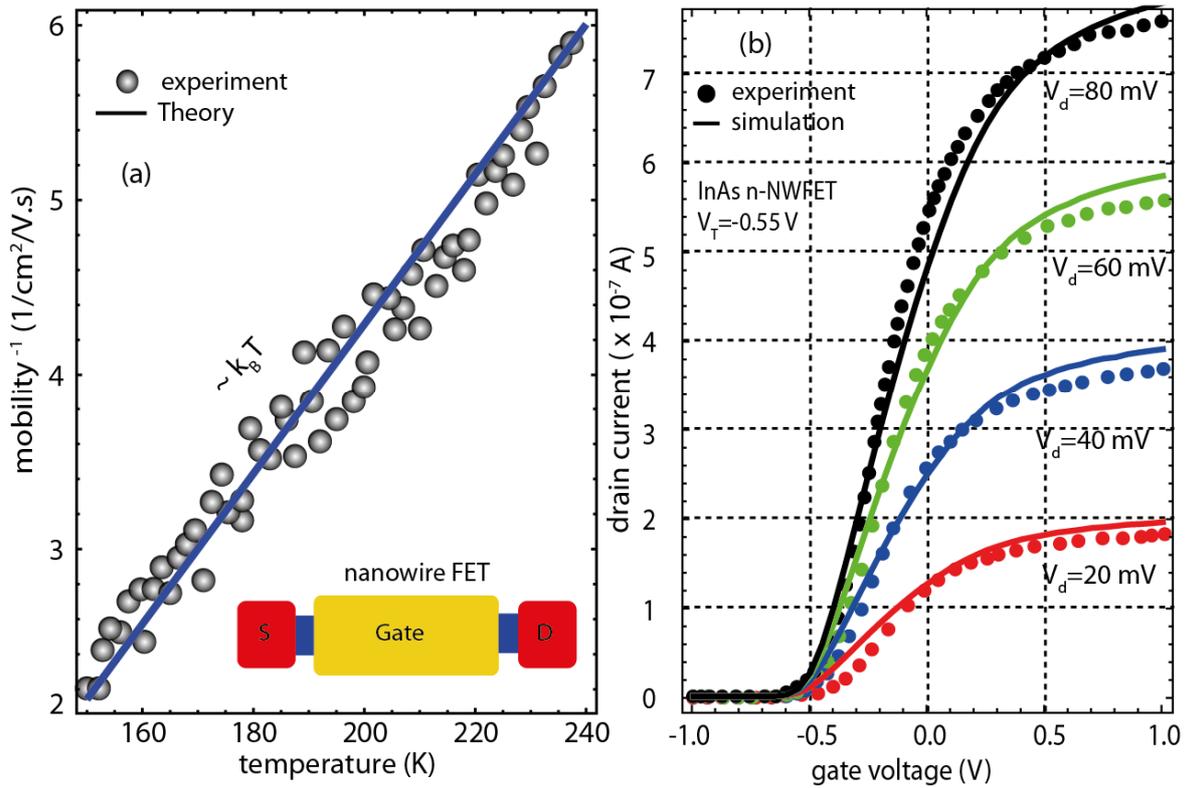

**Figure 5.** Understanding the InAs gate all around FET response: (a) fit of the high-temperature mobility data with phonon scattering mechanism. Extracted sound velocity is found to be almost an order lower due to extended twin boundaries in our InAs nanowires. (b) Comparison of simulated InAs depletion n-NWFET transfer characteristics with experimental data. We use impurity and phonon scattering model in our in-house 3D self-consistent drift-diffusion (with quantum correction) device simulator for transfer characteristics of InAs nanowires. An excellent agreement is found between simulation results and measurement data corroborating our scattering models for diffusive transport developed in this work.

SUPPLIMENTARY INFORMATION:

# Carrier Transport in High Mobility InAs Nanowire Junctionless Transistors

*Aniruddha Konar,*[*,†] *John Mathew*[⊥]*, Kaushik. Nayak,*[*,‡]*, Mohit. Bajaj,* [*,†] *,Rajan K. Pandey, Sajal Dhara*[⊥]*, K. V. R. M. Murali*[†‡]*, Mandar Deshmukh*[⊥]

[†]IBM Semiconductor Research and Development Center, Manyata Embassy Business Park, Nagawara, Bangalore – 560 045, India

[‡]Department of Electrical Engineering, Indian Institute of Technology Bombay, Powai, Mumbai – 400 076, India

[⊥] Department of Condensed Matter Physics and Materials Science, Tata Institute of Fundamental Research, 1 Homi Bhabha Road, Mumbai 400005

[*]E-mail: (A. K.) anikonar@in.ibm.com

The energy dispersion of carrier gas is given by $E_{n,k} = E_n + \frac{\hbar^2 k^2}{2m^*}$, where $E_n$ is the sub-band energy, $\hbar$ is the reduced Planck constant, and k is the carrier wave vector. The corresponding wave-function is given by $|n,k\rangle = \frac{1}{\sqrt{A}} \phi_n(z) e^{ik.r}$, where A is the normalization area and $\phi_n(z)$ is the envelope function. To calculate impurity-limited scattering time for these thick nanowires, we assume a fixed charge of +e located at a nanowire/$HfO_2$ interface. The electrostatic potential in the channel at a position **r**= (r,z) created by this impurity is given by $V(r,z) = \frac{e}{4\pi\varepsilon_0 \kappa_s \sqrt{r^2 + z^2}}$, where $\varepsilon_0$ free space permittivity and $\kappa_s$ is the dielectric constant of the semiconductor. So the perturbation Hamiltonian for the carriers is gi[‡]ven by $H_{imp} = -eV(r,z)$. The corresponding unscreened matrix element of scattering from state $|n,k\rangle$ to another state $|n,k'\rangle$ is given by

---

[‡] Currently with GlobalFoundaries

$$M(q) = \langle n,k' | H_{imp} | n,k \rangle = \frac{e^2}{2\varepsilon_0 \kappa_s q} \left(\frac{b}{b+q}\right)^3, \text{ where } q = |k'-k| \text{ is the change of momentum in}$$

scattering. In this calculation we assumed a Fang-Howard type wave-function[1], i.e. $\phi_n(z) = \frac{b^3}{2} z e^{-bz}$, where $b = \left(\frac{12 m^* e^2}{\varepsilon_0 \kappa_s \hbar^2}\right) \left[N_{dep} + \frac{11}{32} n\right]^{1/3}$, and $N_{dep}$ is the depletion charge density

In the channel. [Note: For evaluating the unscreened matrix element we use the identity[2] $\int_0^\pi e^{iqr\cos\theta} d\theta = 2\pi J_0(qr)$, $J_0(qr)$ is the Bessel function of order zero]. This bare Coulomb potential will be screened by the free carriers in the channel. The screened matrix element is given by $M^{sc}(q) = M(q)/\varepsilon_{2D}(q)$, where $\varepsilon_{2D}(q) = 1 + \frac{G(q) q_{TF}}{q}$, is the two-dimensional Thomas-Fermi screening function. The form factor $G(q)$[3] comes from the quasi-two dimensional nature (finite extent in z-direction, for a prefect 2D gas $G(q)=1$ ) of the carrier gas and $q_{TF} = \frac{m^* e^2}{2\varepsilon_0 \kappa_s \hbar^2}$ is the two-dimensional Thomas-Fermi[4,5] wave vector. With this screened matrix element, the energy-dependent momentum relaxation time is calculated as $\tau(q)^{-1} = (2\pi)^{-2} \int d^2q M^{sc}(q)(1-\cos\theta) d\theta$, where $\theta$ is the angle of scattering. Since, impurity scattering is elastic in nature ($|k| = |k'|$), hence $q = 2k \sin(\theta/2)$, the momentum relaxation time can be written as $\tau^{-1}(k) = n_{imp}^{2D} \left(\frac{e^2}{2\varepsilon_0 \kappa_s}\right)^2 \left(\frac{m^*}{\pi \hbar^3 k^2}\right) I(k)$, where is the dimensionless integral $I(k)$ is defined as $I(k) = \int_0^1 \frac{u^2 P_0(u,k)}{\left[u + G(u,k)\frac{q_{TF}}{2k}\right]^2 \sqrt{1-u^2}} du$. This dimensionless integral $I(k)$ is a weak function of $k$. In charge transport, only scattering rate at the Fermi level is important. So the scattering time has the dependence on Fermi wave-vector as $\tau(k_f) \sim k_f^2$.

. The electron-acoustic phonon scattering in confined semiconducting layer has been investigated by J. Price[6]. Here we assume a Fang-Howard quantum wavefunction and re-derive the electron-acoustic phonon limited mobility. In the equi-partition regime, the acoustic phonon-limited mobility is given by[6] $\mu_{ac} = \left(\frac{e \rho_m v_s^2 \hbar^3}{m^{*2} D_a^2}\right) \frac{F_{ij}}{k_B T}$, where $F_{ij} = \left(\int_0^\infty \phi_i(z) \phi_j(z) dz\right)^2$ is the form-factor, $\rho_m$ is the mass density of InAs, $v_s$ is the sound velocity in InAs and $D_a$ is the acoustic deformation potential. In lowest sub-band occupation limit, the form factor is evaluated as $F_{11} = \frac{b^6}{4} \int_0^\infty z^4 e^{-2bz} dz = 1/3b$. Since the donor density $n_0 << n(V_g)$ is much less than the electrostatically induced carrier density, expanding the exponential, we can write $n \approx n(V_g) + n_0 - \frac{n_0 E_a}{k_B T}$. Using this expansion in the mobility expression, the inverse acoustic-phonon mobility at high-temperatures takes the form $\mu_a^{-1}(T) = A*T - B$, where $A = \left(\frac{m^{*2} D_a^2}{e \rho_m v_s^2 \hbar^3}\right) \left(\frac{12 e^2 m^* N}{\hbar^2 \varepsilon_0 \kappa_s}\right)^{1/3} k_B$ and the other parameter is $B = A \times \frac{11}{96} \left(\frac{n_0 E_a}{N}\right)$, where we define $N \approx n(V_g) + N_{dep}$. We use this expression the main text to explain the temperature-dependent mobility data.

The 3-D numerical device simulations were carried out using FIELDAY. FIELDAY solves a system of coupled partial differential equation governing electrostatics and current transport in a typical semiconductor device i.e. BJT, FETs etc. in drift-diffusion simulation framework. The following partial differential equations for potential distribution and carrier continuity equations for electrons and holes are solved using control volume method in FIELDAY:

$$\nabla \bullet \varepsilon \nabla \phi + q(p - n + N) = 0 \dots\dots\dots\dots(1)$$

$$\frac{1}{q} \nabla . J_n + R_n = \frac{\partial n}{\partial t} \dots\dots\dots\dots\dots\dots(2)$$

$$\frac{1}{q} \nabla . J_p + R_p = \frac{\partial p}{\partial t} \dots\dots\dots\dots\dots\dots(3)$$

In above equations, $n$ and $p$ are electron and hole concentrations, respectively, $\phi$ is electric potential, $\varepsilon$ is dielectric constant of the material, $R_n$ and $R_p$ are the net generation rate of electron and holes, q is the electronic charge and N is net impurity density. The current densities $J_n$, $J_p$ are given by:

$$J_n = q \mu_n n \nabla \left[ \phi + \chi + \frac{E_g}{q} - \frac{\kappa_b T}{q} \ln(\gamma_n) - \frac{\kappa_b T}{q} \ln\left(\frac{N_c}{N_{cr}}\right) + \varphi_{QM} \right] - \kappa_b T \mu_n \nabla n$$

$$J_p = q \mu_p p \nabla \left[ \phi + \chi + \frac{\kappa_b T}{q} \ln(\gamma_p) + \frac{\kappa_b T}{q} \ln\left(\frac{N_v}{N_{vr}}\right) + \varphi_{QM} \right] - \kappa_b T \mu_p \nabla p$$

Here, $\mu_n$, $\mu_p$ are electron and hole mobilities, respectively, $\chi$ is the electron affinity, $E_g$ is the band gap of material, T is the lattice temperature, $\kappa_b$ is the Boltzmann constant, $\gamma_n$, $\gamma_p$ are the gamma functions in the Fermi-Dirac integral for electron and holes, respectively, $N_c$, $N_v$ are the effective density of states (DOS) in conduction and valance band, respectively, $N_{cr}$, $N_{vr}$ are the reference effective density of states (DOS) in conduction and valance band, respectively, and $\varphi_{QM}$ is the quantum potential.

The band-structure parameters $\chi$, $E_g$, $\gamma_n$, $\gamma_p$, $N_c$, $N_v$, $\varphi_{QM}$ vary spatially due to heavy doping and strong 2-dimensional quantum confinement (QC) effects. The above band structure parameters in equations (1) and (2), are evaluated via the quantum correction model in the room temperature drift-diffusion numerical simulation framework. The 2-D QC effects are taken care by calibrating the quantum correction model parameters with the results obtained from the self-consistent Schrodinger-Poisson solver. The quantum correction model treats the extra gradient due to quantum corrected electron density through $\varphi_{QM}$, evaluated at the device grid points after each Newton iteration. The carrier distribution is considered to obey Fermi-Dirac statistics. The physical models employed in the simulation are; Band-gap narrowing due to heavy doping, Normal field effects at the gate oxide interface in the FET inversion layer, Quantum corrections of inversion charge. The recombination terms include the conventional Shockley-Read-Hall expression for recombination via traps, three-particle (Auger) effects, surface recombination, and recombination via trap-assisted and band-to-band tunneling.

The carrier scattering in the semi-classical drift-diffusion transport formulation is treated by the physics based carrier mobility model, which traces its lineage to Mujtaba[7] and Lombardi[8]. The mobility model consists of a low-field mobility model consistent with the Matthiessen's rule and a field-dependent mobility model with empirical high-field corrections. The model takes into account electron-phonon scattering, surface roughness scattering, impurity scattering with coulombic screening and un-screening effects. The electric-field dependent carrier mobility model used in the current density formulations is given by;

$$\mu(F_\parallel, T, N, n, p, E_t) = \frac{2\mu_t}{1 + \left[1 + 4(\mu_t F_\parallel / V_s)^2\right]^{1/2}}$$

$$\mu_t(T, N, n, p, E_t) = \left[\frac{1}{\mu_{ph}} + \frac{1}{\mu_{SR}} + \frac{1}{\mu_C} + \frac{1}{\mu_{HiK}}\right]^{-1}$$

Here, $F_\parallel$ is the Driving force due to parallel electric field; $E_t$ is the Local Transverse electric field; $V_s$ is carrier saturation velocity; $\mu_t$ is the Low-field mobility; $\mu_{ph}$ is the Phonon limited mobility, $\mu_{SR}$ Surface-roughness limited mobility, $\mu_C$ Coulombic scattering limited mobility; $\mu_{HiK}$ mobility limited by presence of remote Coulombic charges at the High-k (HfO$_2$) dielectrics material..

The partial-differential equations (PDE's) (1) – (3) are discretized to obtain a set of nonlinear equations using control volume method. For the spatial discretization, a 3-D mesh consisting of a mixture of prisms and tetrahedra is used. Further, a fully-coupled Newton scheme is used to produce a large sparse linear system of equations from the set of nonlinear equations. A sparse matrix linear solver program is used to obtain the solution of the linear system of equations

**TEM imaging of nanowires**:

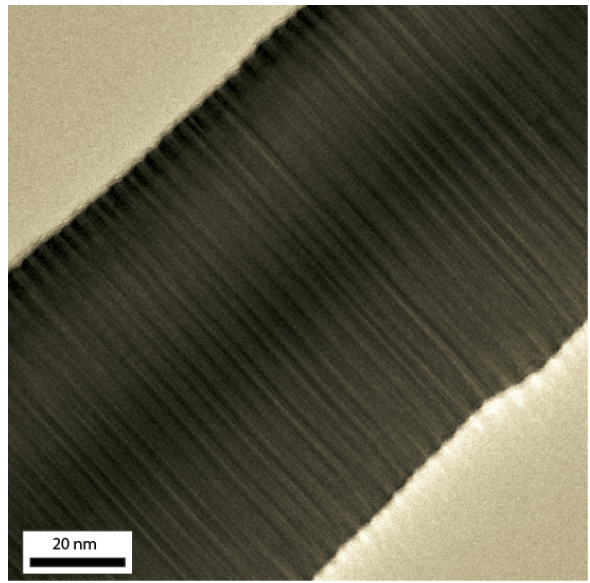

Figure S1 : Transmission electron microscope image of InAs nanowires showing the twin boundaries that result in reduction in acoustic phonon velocity.

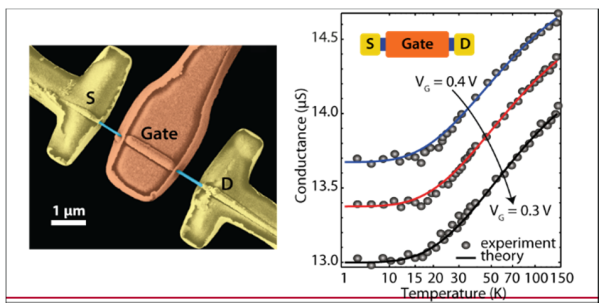